\documentclass[12pt]{article}
\pdfoutput=1
\usepackage{pstricks,slashed}
\usepackage{color,verbatim}
\usepackage{amssymb,amsmath,bm,bbm,accents}
\usepackage{epsf,epsfig}
\usepackage{afterpage}
\usepackage{longtable}
\usepackage{cite}
\usepackage{latexsym,mathrsfs,dsfont}
\usepackage{graphics,graphicx}
\graphicspath{{./Figures/}}
\usepackage{url}
\usepackage{paralist}
\usepackage{bbold}

\usepackage[T1]{fontenc}
\usepackage[utf8]{inputenc}
\usepackage[english]{babel}
\usepackage{braket}
\usepackage{amsthm}
\usepackage{hyperref}
\hypersetup{
    colorlinks=true,
    linkcolor=red,
    filecolor=magenta,      
    urlcolor=cyan,
    }
\usepackage{amsfonts}
\usepackage{tensor}
\usepackage{float}

\newcommand{\be}{\begin{equation}}
\newcommand{\ee}{\end{equation}}
\newcommand{\bi}{\begin{itemize}}
\newcommand{\ei}{\end{itemize}}
\newcommand{\ba}{\begin{array}}
\newcommand{\ea}{\end{array}}
\newcommand{\bea}{\begin{eqnarray}}
\newcommand{\eea}{\end{eqnarray}}
\newcommand{\dd}{\displaystyle}
\newcommand{\nn}{\nonumber}

\newcommand{\azione}{\mathcal{S}}
\newcommand{\lagrangiana}{\mathcal{L}}

\DeclareRobustCommand{\vect}[1]{
  \ifcat#1\relax
    \boldsymbol{#1}
  \else
    \mathbf{#1}
  \fi}
  
\begin{document}

\medskip

\begin{center}
{\Large  Baryon density and magnetic field effects on chaos  \\ \vskip 0.3cm in a $Q \bar Q$ system at finite temperature}
\\[1.0 cm]
{ {Nicola~Losacco}
 \\[0.5 cm]}
{\small 
Istituto Nazionale di Fisica Nucleare, Sezione di Bari,  Via Orabona 4, I-70126 Bari, Italy \\[0.1 cm]
Dipartimento Interateneo di Fisica ``M. Merlin'', Universit\`a  e Politecnico di Bari, \\ via Orabona 4, 70126 Bari, Italy
}
\end{center}

\vskip 0.8cm

\begin{abstract}
\noindent
Baryon density and magnetic field effect on chaos for the holographic dual of a $Q \bar Q$ system at finite temperature is studied. A string in an AdS Reissner–Nordstrom background, and in a metric with magnetic field near the black-hole horizon is considered and small time-dependent perturbations of the static configurations are investigated. The proximity to the horizon induces chaos, which is softened increasing the chemical potential or the magnetic field. A background geometry including the effect of a dilaton is also examined. The Maldacena, Shenker, and Stanford bound on the Lyapunov exponents characterizing the perturbations is satisfied for finite baryon chemical potential and magnetic field and when the dilaton is included in the metric.
\end{abstract}


\setcounter{page}{1}

\section{Introduction}\label{introduction}
The aim of these studies \cite{Colangelo:2020tpr, Colangelo:2021kmn} is to analize the effects of baryon density and magnetic field on the chaotic behavior of a suspended string in a gravitational background.
The works follow the tests, carried out using holographic methods, of the Maldacena-Shenker-Stanford (MSS) bound \cite{Maldacena:2015waa}. The bound, conjectured to be universal, states that, under general conditions, for a thermal quantum system  at temperature $T$ some out-of-time-ordered correlation functions involving  Hermitian operators which characterize the system have an exponential time dependence in determined time intervals.  The dependence is  characterized by the exponent  $\lambda$, for which a bound  (in units where  $\hbar=1$ and $k_B=1$) can be obtained:
\begin{equation}
\lambda \leqslant 2 \pi T .
\label{eq:1}
\end{equation}
The correlation functions quantify quantum chaos. They are the thermal expectation values of the squared commutator of two Hermitian operators at a time separation $t$, that allow to determine the effect of one operator on measurements of the other one at a later time. 

The MSS bound should be satisfied by a set of systems called fast ``scramblers''. The possibility to apply holographic methods to test the bound is supported by the observation that in nature the black holes (BH) are the fastest scramblers: the time needed for a system near a BH horizon to loose information depends  logarithmically on the number of the system degrees of freedom  \cite{Sekino_2008, susskind2011addendum}. Connections between chaotic quantum systems and gravity  have been investigated in  \cite{Shenker_2014,Shenker:2014cwa,Kitaev,Polchinski:2015cea,Giataganas:2021ghs}. In a holographic framework,  a relation has  been worked out  between the size of the operators of the quantum theory on the boundary, which are involved in the temporal evolution of the perturbation, and the momentum of a particle falling in the bulk \cite{Susskind:2018tei,Brown:2018kvn}.

To test the MSS bound \eqref{eq:1} through holographic methods, the quantum system is conjectured to be a $4d$ boundary theory dual to an AdS$_5$ gravity theory with a black hole \cite{Maldacena_1998, Witten:1998qj, Gubser_1998}. 
Several investigations are described in \cite{deBoer:2017xdk,Dalui:2018qqv,Dalui:2020qpt,Ageev:2021xgy,Ma:2022tvs}.
Most of the studies analyze the dynamics of a string hanging in the bulk with endpoints on the boundary, which is the  holographic dual of a static quark-antiquark pair  \cite{Avramis:2006nv,Arias_2010,Nunez:2009da,Bellantuono:2017msk}.
To quantify the chaotic dynamics of such systems, the Lyapunov exponent $\lambda$, characterizing the chaotic behavior of the  fluctuations around the static string configuration is evaluated \cite{Hashimoto:2018fkb, Ishii_2017,Akutagawa_2019}. In the work \cite{Halder:2019ric} the bound has been generalized to a quantum system in a thermal ensamble and a global symmetry. In the case of QCD, the global symmetry can be related to the conservation of the baryon number. The generalization in \cite{Halder:2019ric} relaxes the MSS bound:
\begin{equation}
\lambda \leqslant \frac{2 \pi T}{1-\left|\frac{\mu}{\mu_c}\right|} ,
\label{eq:2}
\end{equation}
\noindent
where $\mu$ is the chemical potential related to the global
symmetry, and $\mu_c$ a critical value above which the
thermodynamic ensemble is not defined. The inequality
\eqref{eq:2} is conjectured for $\mu \ll \mu_c$. This means that systems that satisfy the bound \eqref{eq:2} could violate the MSS one.

In this review a general approach to study the chaotic behavior of such systems is presented.
Two applications of the procedure are analyzed. The first one aims to test the generalized bound, considering the role of a $U(1)$ global symmetry connected to the conservation of the baryon number. In such case, a dual metric can be identified with the AdS Reissner-Nordstrom (RN) metric with a charged black hole. We can use this background for testing Eq.\eqref{eq:2}.
The second case analyzes the impact of an external magnetic field on the chaotic behaviour of the string. The magnetic field is relevant in different contexts, including  heavy-ion collisions or condensed matter problems such as the Quantum Hall Effect and superconductivity at high temperatures. A general gravity dual for such systems is presented in \cite{Critelli:2016cvq,Ballon-Bayona:2020xtf,Arefeva:2020vae,Arefeva:2020bjk,Arefeva:2021mag,Arefeva:2022avn}.
The backreaction of an external magnetic field modifies the geometry of the $5d$ spacetime, the metric of which is determined by the Einstein equations. As a result, an anisotropy  is  introduced in the spatial directions. 
Moreover, in a finite temperature system the relation between the position of the black hole horizon, the source of chaos in the $5d$ geometry, and temperature, involved in the MSS relation in the boundary theory, is modified by the magnetic field.

\section{String profile in the gravitational background}\label{profile}

\begin{figure}[h]
    \centering
    \includegraphics[width=0.5 \textwidth]{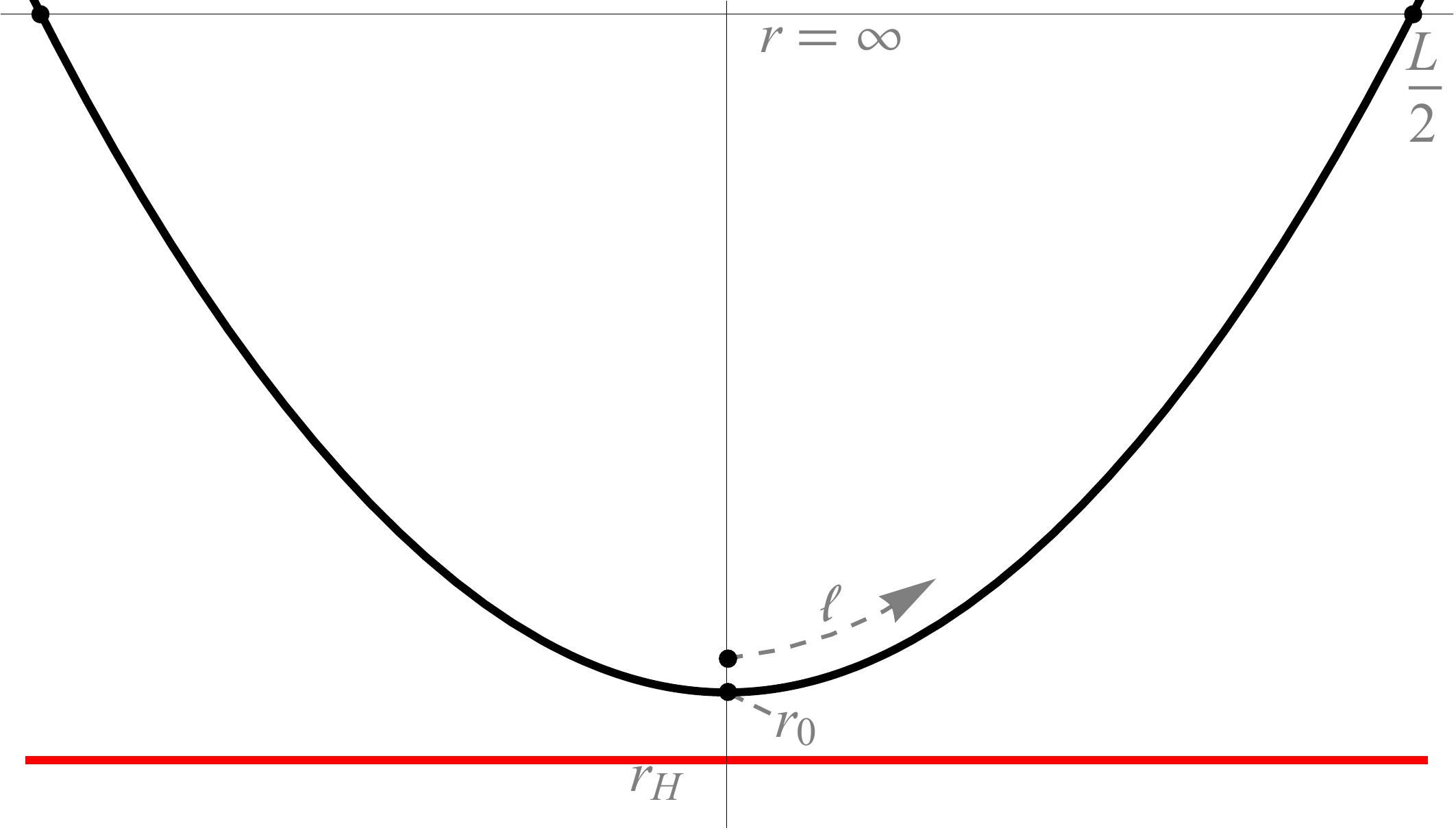}
    \caption{\small Profile of a static string for the $Q \bar Q$ system. $r_0$ is the position of the tip of the string, $r_H$ the position of the horizon, $L$ the distance between the end points on the boundary.}
    \label{Fig:1}
\end{figure}

We are interested in the gravity dual of a strongly coupled $Q \bar Q$ pair in a general thermodynamic background at finite temperature. The hanging string is described by the functions $r(t,\ell)$ and $x_i(t,\ell)$, in an asymptotically AdS$_5$ geometry with a black hole. The endpoints of the string are on the AdS boundary $r\to\infty$.
$(t,\ell)$ are the worldsheet coordinates, with $\ell$  the proper distance measured along the string, Fig. \ref{Fig:1}.

\noindent
The line element for a generic $5$ dimensional diagonal metric can be expressed as

\begin{equation}\label{line}
    ds^2=g_{tt} dt^2 + g_{11}(dx^1)^2+ g_{22} (dx^2)^2 +g_{33} (dx^3)^2+g_{rr} dr^2.
\end{equation}
\noindent
The string dynamics is governed by the Nambu-Goto (NG)  action:
\begin{equation}
    \azione=-\frac{1}{2\pi \alpha^\prime} \int dt\, d \ell \, \sqrt{-h}\,,
\end{equation}
where $\alpha^\prime$ is the string tension and $h$ the determinant of the induced metric $h_{ij}=g_{MN} \frac{\partial X^M}{\partial \xi_i} \frac{\partial X^N}{\partial \xi_j}$, with $\xi_{i,j}$ the worldsheet coordinates and $g$ the metric tensor.
In the static case the action reads:
\begin{equation}
    \azione = -\frac{T}{2\pi\alpha^\prime}\int d\ell \, \sqrt{|g_{tt} g_{ii} {\acute x_i}^2 + g_{tt} g_{rr} {\acute r}^2|} \,,
\end{equation}
where $\acute x_i$ denotes the derivative with respect to $\ell$.
$x_i$ is a cyclic coordinate, so its conjugate momentum
\begin{equation}
\frac{\partial \lagrangiana}{\partial \acute x_i}=-\frac{T}{2\pi\alpha^\prime} \frac{|g_{tt}| g_{ii} {\acute x_i}}{\sqrt{|g_{tt}| g_{ii} {\acute x_i}^2 + |g_{tt}| g_{rr} {\acute r}^2}}
\end{equation}
is a constant of motion.
Denoting with $r(\ell=0) = r_0$ the position of the tip of the string in the bulk, i.e. the point where $\dd \frac{dr}{dx_i}\Big |_{\ell=0}=0$, we have:
\begin{eqnarray}
\frac{\sqrt{|g_{tt}|} g_{ii} {\acute x_i}}{\sqrt{g_{ii} {\acute x_i}^2 + g_{rr} {\acute r}^2}} =\left. \sqrt{|g_{tt}| g_{ii}} \right|_{\ell=0}\,.
\end{eqnarray}
Moreover, from the condition
\begin{eqnarray}
&& d\ell^2=g_{ii}\, dx_i^2+g_{rr}\, dr^2 
\end{eqnarray}
the equations determining the string profile can be obtained:
\begin{eqnarray}\label{eq:eqxstatic}
\acute x &=& \pm \frac{\sqrt{-g_{tt}(r_0) g_{ii}(r_0)}}{\sqrt{-g_{tt}} g_{ii}} \\
\acute r &=& \pm \frac{\sqrt{-g_{tt} g_{ii} + g_{tt}(r_0) g_{ii}(r_0) }}{\sqrt{-g_{tt} g_{ii} g_{rr}}}\,.
\label{eq:eqrstatic}
\end{eqnarray}
We set the string endpoints lying on the AdS$_5$ boundary at $x_i=\pm L/2$. The minimum value $r_0$ of the coordinate $r$ is reached at $x_i=0$ (or $\ell = 0$). $L$ and $r_0$  are related, since
\begin{equation}\label{eq:Lvsr0}
    L = 2 \int_{r_0}^\infty dr \, \Bigg(\frac{g_{ii}(r)}{g_{rr}(r)} \left( \frac{g_{tt}(r) g_{ii}(r)}{g_{tt}(r_0) g_{ii}(r_0)}-1\right) \Bigg)^{-\frac{1}{2}} \,.
\end{equation}
Therefore, the static string configuration depends on $r_0$ or $L$. Choosing different configurations, it is possible to probe the effect of the closness of the BH horizon on the chaotic behaviour of the string. As shown in Ref.~\cite{Colangelo:2020tpr} the proximity to the horizon enhances the chaotic behaviour, hence we choose an  unstable configuration near the horizon as starting point to perturb and study the dynamics of the fluctuation.

\section{Perturbing the static solution}\label{expansion}
The chaotic dynamics can be studied by perturbing the static string configuration near the black hole horizon.
\begin{figure}[b!]
	\centering
	\includegraphics[width=0.5 \textwidth]{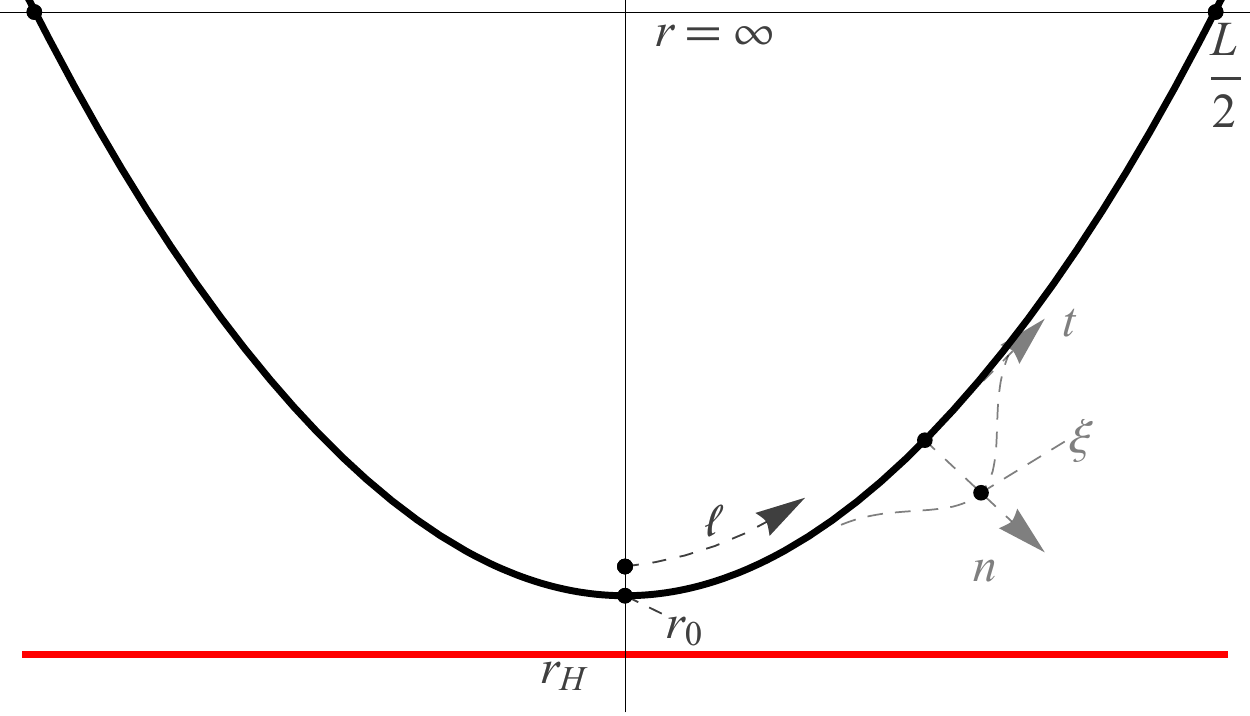}
	\caption[\small Perturbed string]{\baselineskip 12 pt \small Static string profile and perturbation $\xi(t,\ell)$ along the direction orthogonal to the string in each point with coordinate $\ell$.}
	\label{Fig:2}
\end{figure}
The perturbation is chosen to be orthogonal in each point of the string, and described by the coordinate $\ell$ in the $r-x$ plane \cite{Hashimoto:2018fkb,Colangelo:2020tpr}.  The perturbation is depicted in Fig.~\ref{Fig:2}.
Considering the unit vector $n^M= (0,n^x,0,0,n^r)$ orthogonal to $t^M$, we have:

\bea
g_{rr} ( r ) \left({n}{^{r}}\right)^2 +g_{xx} ( r) \left({n}{^{x}}\right)^2 &=& 1
\label{eq:1M} \\
\acute{r} \left( \ell \right)  g_{rr} \left( r \right)  \, n^r + \acute{x} \left( \ell \right)  g_{xx}\left( r \right)  \, n^x&=&0 \,.
\label{eq:2M}
\eea
For an outward perturbation as  in Fig.~\ref{Fig:2}  the solution for the components $n^{x}$ and $n^{r}$  is

\begin{equation}
{n^x} ( \ell)=\sqrt{\frac{g_{rr}}{g_{xx}}} \; \acute{r}( \ell ) \quad \text{,} \quad {n}{^{r}} ( \ell )=-\sqrt{\frac{g_{xx}}{g_{rr}}} \; \acute{x}( \ell ) \quad .
\label{eq:3}
\end{equation}
\noindent 
The time-dependent perturbation $\xi \left( t, \ell \right)$  modifies $r$ and $x$:
\bea
r \left( t,\ell \right) &=& r_{BG} \left( \ell \right) + \xi \left( t, \ell \right) n^{r} \left( \ell \right), \nn \\
x\left( t,\ell \right) &= & x_{BG} \left( \ell \right) + \xi \left( t,\ell \right) n^{x} \left( \ell \right),
\label{eq:4} 
\eea
where $r_{BG} \left( \ell \right)$ and $x_{BG} \left( \ell \right)$  are the static solutions obtained integrating Eqs.~\eqref{eq:eqxstatic} and \eqref{eq:eqrstatic}. 

The dynamics of the small perturbation can be analyzed expanding the metric function around the static solution $r_{BG}(\ell)$  to the third order in  $\xi \left( t,\ell \right)$.
To this order the NG action comprises a quadratic and a cubic term.
The  quadratic term  has the form:

\begin{equation}
\begin{aligned}
S^{\left( 2 \right)} = \frac{1}{2 \pi \alpha^\prime}\int \mathrm{d}t \int_{-\infty}^{\infty} \mathrm{d} \ell \left( C_{tt} \dot{\xi}^2 + C_{\ell \ell}  \acute{\xi}^2  + C_{00} \xi^2\right).
\end{aligned}
\label{eq:2nd Order Action}
\end{equation}
\noindent
$C_{tt}$, $C_{\ell \ell}$ and $C_{00}$ depend on $\ell$ and on the parameters of the metric. 
The  equation of motion from the action \eqref{eq:2nd Order Action} is
\begin{equation}
\qquad \qquad C_{tt} \, \ddot{\xi} + \partial_\ell \left( C_{\ell \ell} \acute{\xi} \right) - C_{00} \, \xi = 0.
\label{eq:8}
\end{equation}
Factorizing $\xi \left( t,\ell \right) = \xi \left(\ell \right) e^{i \omega t}$ it  corresponds to the Sturm-Liouville equation 
\be
\partial_\ell \left( C_{\ell \ell} \, \acute{\xi} \right) - C_{00} \, \xi = \omega ^2 C_{tt} \, \xi \,\, ,
\label{eq:9}
\ee
\noindent
with $W(\ell)=-C_{tt}(\ell)$ the weight function.
Eq.~\eqref{eq:9} can be solved for different values of the parameters characterizing the metric. The  two lowest  eigenvalues $\omega_0^2$ and $\omega_1^2$, with the corresponding eigenfunctions $\xi \left(\ell \right)=e_0 \left(\ell \right)$ and $\xi \left(\ell \right)=e_1 \left(\ell \right)$ can be obtained.
The third order terms in $\xi$ in the action give us information on the chaotic behaviour. 
Up to a surface term, the  expression  is

\be
S^{\left( 3 \right)} =  \frac{1}{2 \pi \alpha^\prime}\int \mathrm{d}t \int_{-\infty}^{\infty} \mathrm{d}\ell \bigg\{ D_0 \,   \xi^3 
+ D_1 \, \xi \acute{\xi}^2 + D_2 \, \xi \dot{\xi}^2  \bigg \} \,\,  ,
\label{eq:11}
\ee
with $D_{0,1,2}$ functions of $\ell$.
Expanding the  perturbation in  terms of the first two eigenfunctions  $e_0(\ell)$ and $e_1(\ell)$,

\begin{equation}
\xi \left( t,\ell \right) = c_0 \left( t \right) e_0 \left( \ell \right) + c_1 \left( t \right) e_1 \left( \ell \right) ,
\label{eq:10}
\end{equation} 
\noindent 
the time dependence of the perturbation is encoded in the coefficients $c_0( t )$ and $c_1 ( t )$. 
With this form of $\xi(t,\ell)$  we have

\bea
S^{\left( 3 \right)}&&=\nn \\
&& \frac{1}{2 \pi \alpha^\prime}\int \mathrm{d}t \int_{-\infty}^{\infty} \mathrm{d}\ell \Big[ \Big( D_0  \,  e_0^3 + D_1 \, e_0 \acute{e}_0^2 \Big) c_0^3 \left( t \right) + \Big( 3 D_0  \, e_0 e_1^2 + D_1  \left( 2 \acute{e}_0 e_1 \acute{e}_1 + e_0 \acute{e}_1^2 \right)\Big) c_0 c_1^2  \nn \\
&&+ D_2  \Big( e_0 e_1^2 c_0 \dot{c}_1^2 +e_0^3 e_1^2 c_0 \dot{c}_0^2+ 2 e_0 e_1^2 \dot{c}_0 c_1 \dot{c}_1    \Big) \Big]. 
\label{eq:12}
\eea
The action for $c_0(t)$ and $c_1(t)$ is obtained by $S^{(2)}+S^{(3)}$, integrating over $\ell$:

\be
S^{(2)}+S^{(3)}=\frac{1}{2 \pi \alpha^\prime}\int \mathrm{d}t \Big[ \sum_{n=0,1}\left(\dot{c}_n^2-\omega_n^2 c_n^2  \right) + K_1 c_0^3 
+ K_2 c_0 c_1^2  + K_3 c_0 \dot{c}_0^2 + K_4 c_0 \dot{c}_1^2 + K_5 \dot{c}_0 c_1 \dot{c}_1\Big].
\label{eq:13}
\ee
\noindent
The  coefficients $K_{1, \dots, 5}$ depend on $r_0$ and on the parameters of the metric.
In general the potential described by Eq. \eqref{eq:13} has a trap for the unstable string configurations. We are interested in the motion of $c_0$ and $c_1$ in the trap. In some regions of the  potential the kinetic term is negative.  As shown in \cite{Hashimoto:2018fkb,Colangelo:2020tpr},  it is useful to replace $c_{0,1}\to \tilde c_{0,1}$ in the action, with $c_0=\tilde{c}_0 + \alpha_1 \tilde{c}_0^2 + \alpha_2 \tilde{c}_1^2$ and $c_1 = \tilde{c}_1 + \alpha_3 \tilde{c}_0 \tilde{c}_1$, neglecting  $\mathcal{O} \left( \tilde{c}_i^4 \right)$ terms, setting the constants $\alpha_i$  ensuring the positivity of the kinetic term. This replacement stretches the potential and stabilizes the time evolution of the system. The dynamics is not affected, and a chaotic behaviour shows up in the transformed system.
\newpage
\section{Geometry}\label{geometry}
The procedure can now be applied to specific cases. Interesting systems could be a $Q \bar Q$ pair in a finite temperature and baryon density background, and the pair in a constant and uniform magnetic field at finite temperature. The first one would allow us to test \eqref{eq:2}, and therefore to relax the MSS bound in the presenece of a global symmetry, and the other is relevant in different phenomenological contexts.
We need to specify the dual metric of the systems
solving Einstein equations with suitable boundary conditions.

\subsection{Finite baryon density}\label{baryon}

Consider a Lagrangian of a quantum field theory with a $U \left( 1 \right)$ gauge symmetry and a Dirac fermion charged under this symmetry,

\begin{equation}
\mathcal{L} = i \bar{\psi} \gamma^{\mu} D_{\mu} \psi - \frac{1}{4 g^2} F^{\mu \nu} F_{\mu \nu},
\label{eq:3.100}
\end{equation}
\noindent with the covariant derivative $D_{\mu} = \partial_{\mu} + i A_{\mu}$. A chemical potential can be introduced considering a non-vanishing background field

\begin{equation}
A_0 = \mu, \qquad A_i = 0 \quad (i = 1, \dots, 3).
\label{eq:3.101}
\end{equation}
\noindent This generates a potential of the form 

\begin{equation}
V= - \mu ~ \psi^{\dagger}\psi .
\end{equation} 
\label{eq:3.102}
\noindent Since $\psi^{\dagger}\psi = \hat{N}$ is the number operator, $\mu$ is the chemical potential by definition: it represents the change in energy of the system when a particle is added.

We are interested in finding the $AdS$ gravity dual of this system. In general the gravity action describing the $5$-dimensional asymptotic AdS space and the gauge field is given by 

\begin{equation}
S = \int d^5 x \sqrt{-G} \left[ \frac{1}{2 \mathit{k}^2} \left( \mathcal{R} - 2 \Lambda \right) - \frac{1}{4 g^2} F_{M N} F^{M N}\right],
\label{eq:3.103}
\end{equation}
\noindent where $2 \mathit{k}^2$ is proportional to the five-dimensional Newton constant and $g^2$ is a five-dimensional gauge coupling constant. In the $ AdS_{5} $ space, the cosmological constant is given by $\Lambda = -6/L^2$, where $L$ is the radius of the $AdS$ space. The equations of motion of this system are

\begin{equation}
\begin{aligned}
&\mathcal{R}_{MN} - \frac{1}{2} G_{M N} \mathcal{R} + G_{M N} \Lambda = \frac{\mathit{k}^2}{g^2} \left( F_{M P} \tensor{F}{^P_N} - \frac{1}{4} G_{M N} F_{P Q} F^{P Q} \right), \\
& 0 = \partial_M \sqrt{-G}~G^{M P} ~ G^{N Q} F_{P Q},
\end{aligned}
\label{eq:3.104}
\end{equation} 
\noindent the Einstein equations and the Maxwell ones. Knowing the five-dimensional gauge field $A^M$ in the $AdS$ space, Eq.~\eqref{eq:3.104} gives the metric of the space. To recover the gauge theory at the boundary of the $AdS$ space we set

\begin{equation}
A_0 \left( x^ {\mu}, z \right) =  \mu - Q z^2, \qquad A_i = A_4 = 0 \quad (i = 1,2,3), 
\label{eq:3.105}
\end{equation}
\noindent where $z= 1/r$ is the bulk coordinate in the Fefferman-Graham coordinate system. The line element of a $5$-dimensional asymtotic $AdS$ spacetime reads

\begin{equation}
ds^2 = \frac{L^2}{z^2} \left( - f \left( z \right) d t^2 + d \vect{x}^2 +\frac{1}{ f \left( z \right)} d z^2 \right),
\label{eq:3.106}
\end{equation} 
\noindent and the boundary is $z=0$.

Solving Eq.~\eqref{eq:3.104} we obtain the Reissner-Nordstrom $AdS$ black-hole

\begin{equation}
f \left( z \right) = 1 - m z^4 + q^2 z^6,
\label{eq:3.107}
\end{equation}
\noindent where $m$ is the mass of the black-hole and $q$ is its charge. Introducing Eq.~\eqref{eq:3.105} and Eq.~\eqref{eq:3.107} in Eq.~\eqref{eq:3.104} and evaluating it a the boundary $z=0$ we obtain a relation between $q$ and $Q$

\begin{equation}
q^2 = \frac{2 \mathit{k}^2}{3 g^2 L^2} Q^2.
\label{eq:3.108}
\end{equation}
\noindent In the $AdS/QCD$ context, the gravitation constant $2 \mathit{k}^2$ and the $5$-dimensional coupling constant $g^2$ are related to the rank of the gauge group $N_c$ and the number of flavors $N_f$ in QCD \cite{Lee_2009}

\begin{equation}
\frac{1}{2 \mathit{k}^2} = \frac{N_c^2}{8 \pi^2 L^3} \quad \text{and} \quad \frac{1}{g^2} = \frac{N_c N_f}{4 \pi^2 L},
\label{eq:3.109}
\end{equation} 
\noindent therefore, Eq.~\eqref{eq:3.109} can be written as

\begin{equation}
q = \sqrt{\frac{2}{3} \frac{N_f}{N_c}} ~ Q.
\label{eq:3.110}
\end{equation}
\noindent
The RN black-hole has two horizons identified by the solutions of 
\begin{equation}
f \left( z_h \right) = 1 - m z_h ^4 +q^2 z_h^6 = 0.
\label{eq:3.111}
\end{equation}
\noindent We can write the mass of the black-hole as a function of the outer horizon and of its charge:

\begin{equation}
m=\frac{1}{z_h^4} + q^2 z_h^2.
\label{eq:3.112}
\end{equation} 
\noindent 
To show the relation between the charge $q$ and the chemical potential $ \mu $ we can observe that to have a regular norm $ A^M \left( x^{\mu}, z \right) A_M \left( x^{\mu}, z \right)  \equiv G^{00} A_0 A_0$, $A^0\left( x^{\mu}, z \right)$ should vanish at the outer horizon therefore

\begin{equation}
A^0\left( x^{\mu}, z_h \right) = \mu - Q z_h^2 = 0.
\label{eq:3.113}
\end{equation}
\noindent This gives

\begin{equation}
Q = \frac{\mu}{z_h^2}.
\label{eq:3.114}
\end{equation}
\noindent Using Eq.~\eqref{eq:3.110} we have the relation between the charge of the black-hole and the chemical potential

\begin{equation}
q = \sqrt{\frac{2}{3} \frac{N_f}{N_c}} \frac{\mu}{z_h^2}.
\label{eq:3.115}
\end{equation}
\noindent Inserting Eq.~\eqref{eq:3.112} and Eq.~\eqref{eq:3.115} in Eq.~\eqref{eq:3.107} we obtain:

\begin{equation}
\begin{aligned}
f \left( z \right) &= 1 - \frac{z^4}{z_h^4} -\left( \frac{2}{3} \frac{N_f}{N_c} \right) \frac{\mu^2 z^4}{z_h^2} + \left( \frac{2}{3} \frac{N_f}{N_c} \right) \frac{z^6 \mu ^2}{z_h^4}\\
&=  1 - \frac{z^4}{z_h^4} - \frac{\mu^2 z^4}{z_h^2} + \frac{z^6 \mu ^2}{z_h^4},
\end{aligned}
\label{eq:3.116}
\end{equation}
\noindent where the coefficient $ \frac{2}{3} \frac{N_f}{N_c}$ is absorbed in the definition of $\mu^2$. Returning to the $r$ coordinates we obtain the RN metric function

\begin{equation}
\begin{aligned}
f \left( r \right) =  1 - \frac{r_h^4}{r^4} - \frac{\mu^2 r_h^2}{r^4} + \frac{r_h^4 \mu ^2}{r^6}.
\end{aligned}
\label{eq:RNMF}
\end{equation}
\noindent
The line element is obtained from Eq.~\eqref{eq:3.106} setting $L=1$

\begin{equation}
d s^2 = - f \left( r \right) r^2 d t^2 + r^2 d \vect{x}^2 +\frac{1}{ r^2 f \left( r \right)} d r^2 ,
\label{eq:RNLine}
\end{equation} 
\noindent
and the Hawking temperature $T$ is

\begin{equation}\label{eq:Tvsrhmu}
    T = \frac{r_h}{\pi} \left|1 - \frac{\mu^2}{2 r_h^2}\right| \,.
\end{equation}

\subsection{Magnetic field}\label{magnetic}

A magnetic field is introduced in the holographic framework by a $U(1)$ gauge field $F_{MN}$ which modifies the $5d$ geometry.
The metric is determined solving the Einstein equations:
\begin{equation}
    R_{MN}-\frac{1}{2} g_{MN} (R+12)-T_{MN}=0 
    \label{eq:einstein}
\end{equation}
with the $5d$ stress-energy tensor
\begin{equation}
    T_{MN}= 2 \, (g^{AB} F_{MA} F_{NB} -\frac{1}{4} g_{MN} F^2) \,.
    \label{eq:stress}
\end{equation}
For a constant magnetic field $B$ in the $x_3$ direction $F$ is given by $F=B \, dx^1\wedge dx^2$, hence the only nonvanishing components are $F_{12}=-F_{21}=B$.
The Einstein equations have been solved perturbatively in the low-$B$ and high temperature limits in Refs.\cite{DHoker:2009mmn,DHoker:2009ixq,Li:2016gfn}. The result for the line element, having the general expression
\begin{equation}\label{eq:metricHuang}
    ds^2=g_{tt} dt^2 + g_{11}(dx^1)^2+ g_{22} (dx^2)^2 +g_{33} (dx^3)^2+g_{rr} dr^2 
\end{equation}
with $r>r_h$,  reads:  
\be \label{eq:metricHuang1}
g_{tt}=-r^2 f(r) , \quad g_{11}=g_{22}=r^2 h(r),  \quad g_{33}=r^2 q(r),  \quad g_{rr}=\frac{1}{r^2 f(r)}. 
\ee
The metric functions are expressed as \cite{Li:2016gfn}
\begin{eqnarray}
    f(r)&=&1-\frac{2B^2}{3 r^4}  \log r+\frac{f_4}{r^4} \label{eq:gtt}\\
    q(r)&=&1-\frac{2 B^2}{3 r^4}  \log r \label{eq:g33} \\
    h(r)&=&1+\frac{B^2}{3 r^4}  \log r\,. \label{eq:g22}
\end{eqnarray}
The magnetic field breaks rotational invariance, hence  $g_{22}\neq g_{33}$.
The geometry has a horizon, the position of which $r_h$  is found  requiring $f(r_h)=0$. This gives $\dd f_4 = - r_h^4 + \frac{2}{3}\, B^2 \log(r_h)$
and the blackening function $f(r)$
\be
f(r)=1-\frac{r_h^4}{r^4}-\frac{2 B^2}{3 r^4}  \log {\frac{r}{r_h}} . \label{eq:gtt1}
 \ee
The  Hawking temperature $T$ depends on the magnetic field: 
\begin{equation}\label{eq:TvsrhB}
    T = \frac{r_h}{\pi} \left(1 - \frac{B^2}{6 r_h^4}\right) \,.
\end{equation}
The metric given in terms of the functions \eqref{eq:gtt}-\eqref{eq:g22}  is  obtained  for large bulk coordinate $r$ and low $B$,  and it is important to reckon the minimum value of $r$ and the largest value of $B$ for which it is a good approximation of Eqs. \eqref{eq:einstein},\eqref{eq:stress} and \eqref{eq:metricHuang1}.
This metric can be compared to the one obtained by numerical solution of the Einstein equation. The comparison shows that the deviation in using the approximated metric are small and can be safely used.

\section{Poincar\'e sections and Lyapunov exponents}\label{modes}
\begin{figure*}[h!]
\centering
\makebox[\linewidth][c]{{
	{\includegraphics[width=0.5 \textwidth]{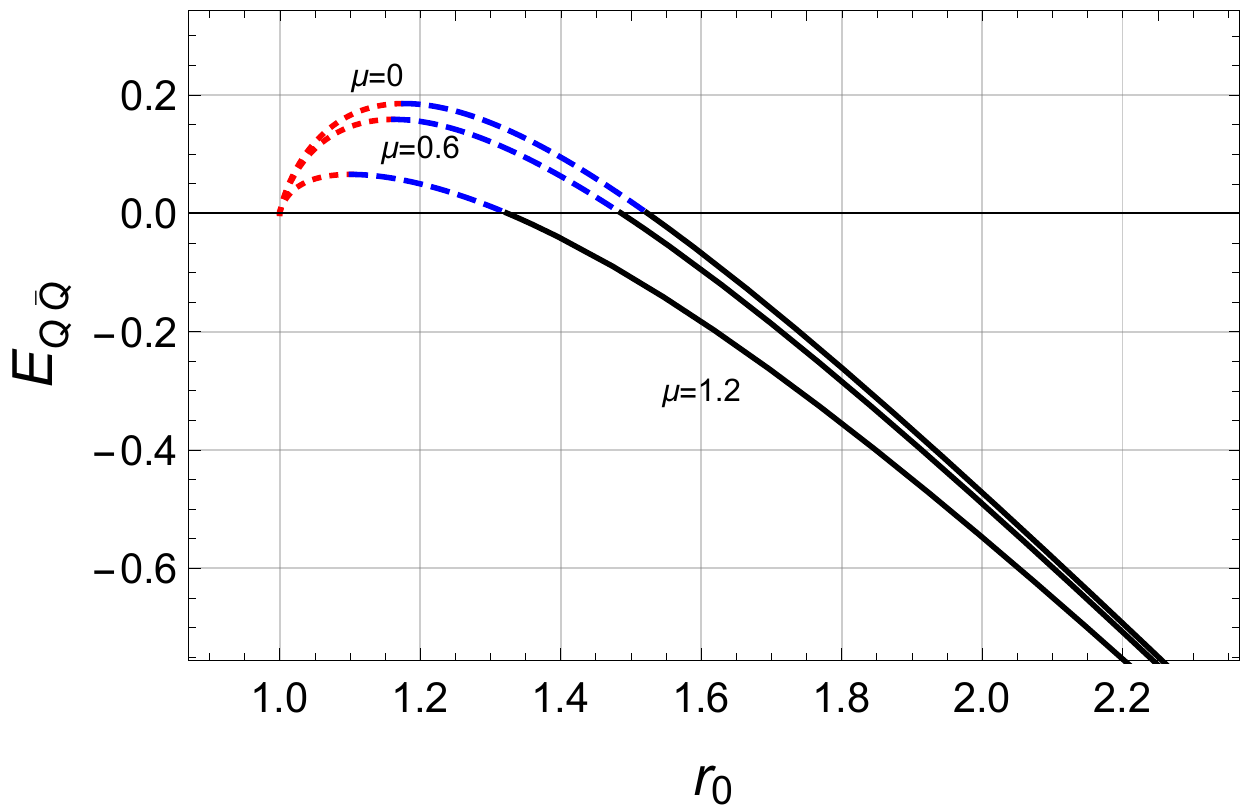} \quad
	 \includegraphics[width=0.5 \textwidth]{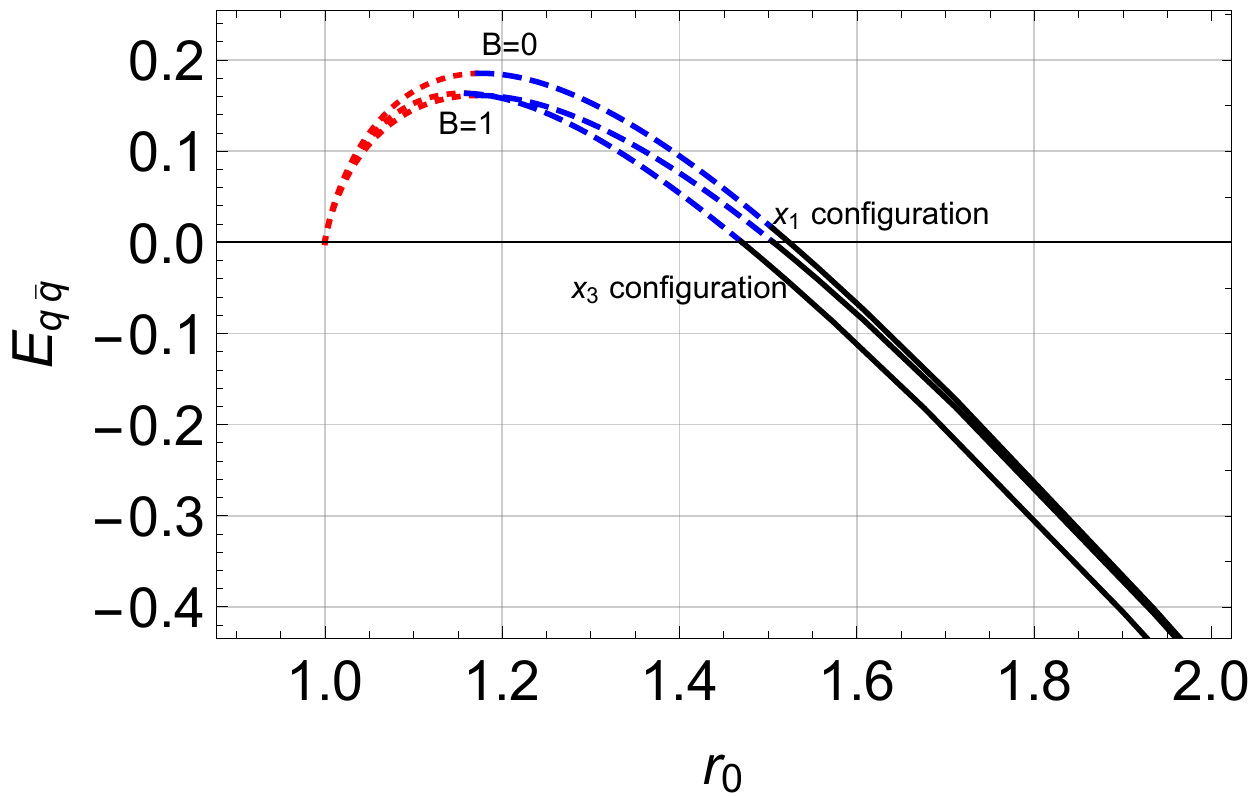}}}} 
\quad
\caption[Energy]{\small Energy of the string configuration as a function of $r_0$ for different values of chemical potential $\mu$ (left) and magnetic field $B$ (right). The maximum divides the unstable configurations (red dotted line) form the metastable (blue dashed line) and stable (solid black line) ones.}
\label{Fig:energy}
\end{figure*}

We can apply the procedure discussed in section \ref{profile} and \ref{expansion}.
We start by choosing a static string configuration fixing the tip position of the string. We consider the energy of the string configuration as a function of $r_0$, Fig.~\ref{Fig:energy}, and observe that the the energy has a peak near the BH horizon. The configurations closer to the horizon are unstable, and the effect of chaos is enanched. $r_0 = 1.1$ is choosen as the tip position for the static configuration for both metrics.

\begin{figure*}[t!]
\centering
\makebox[\linewidth][c]{{
	{\includegraphics[width=0.4 \textwidth]{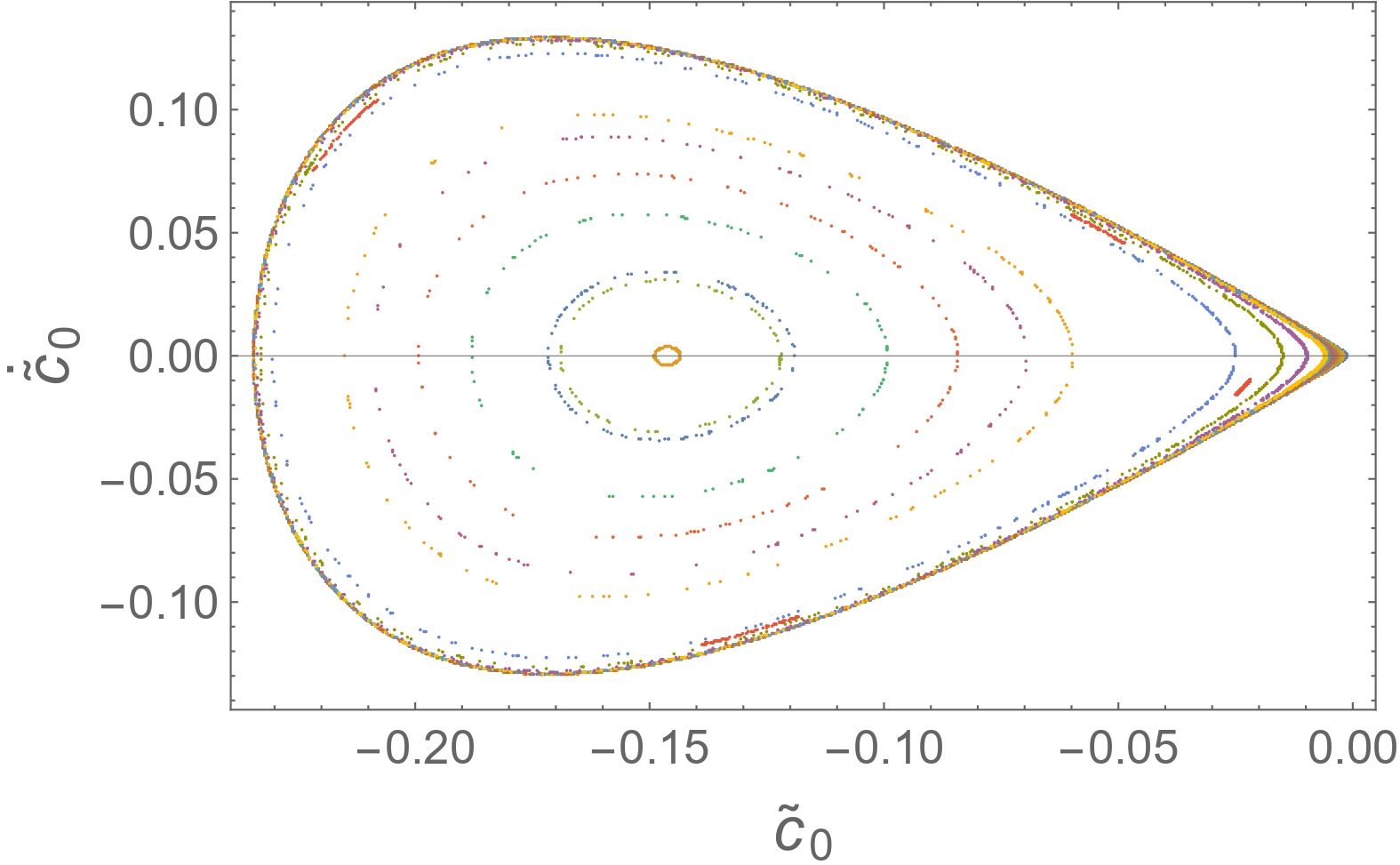} \quad
	 \includegraphics[width=0.4 \textwidth]{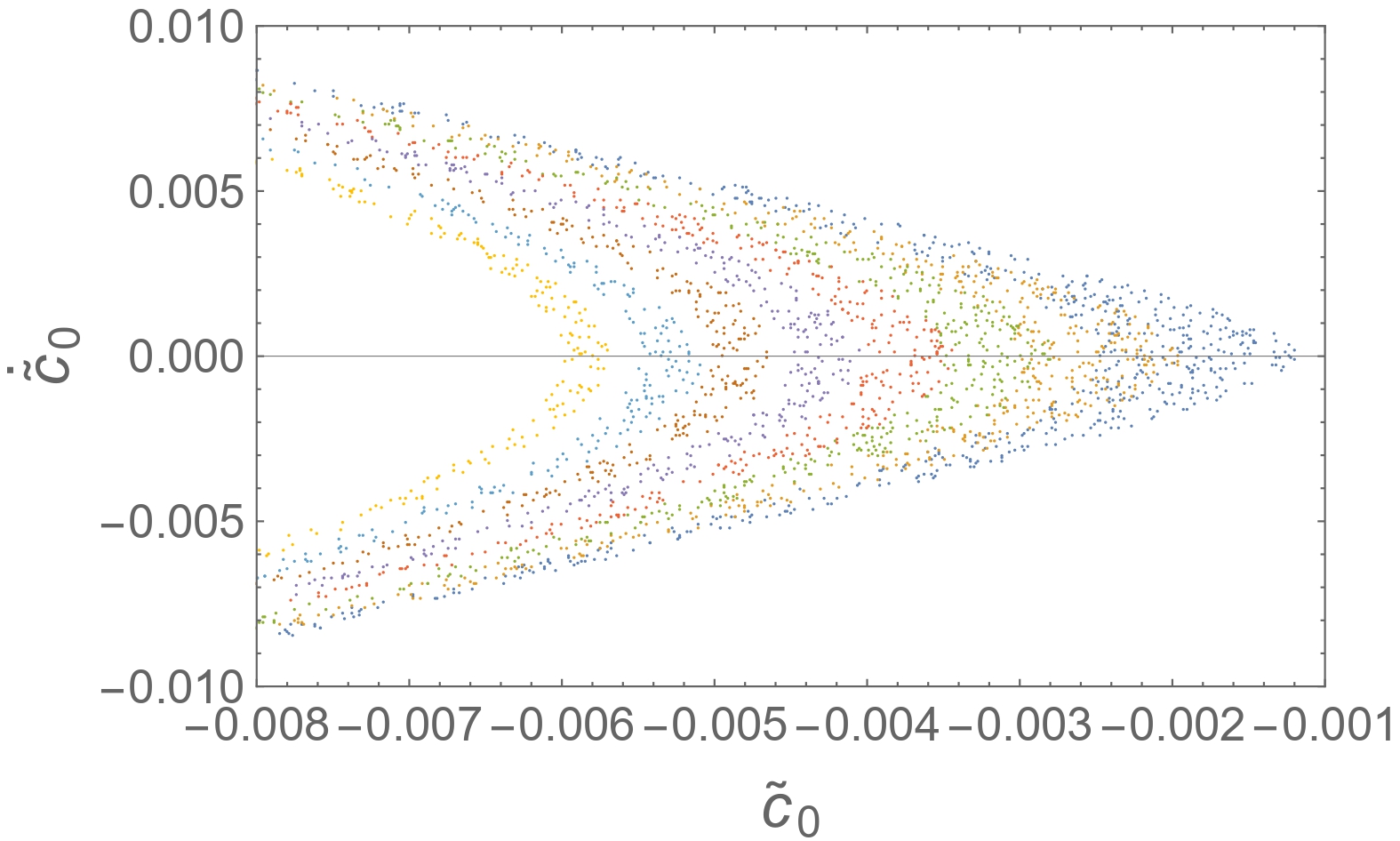}}}} \quad
\makebox[\linewidth][c]{{
	{\includegraphics[width=0.4 \textwidth]{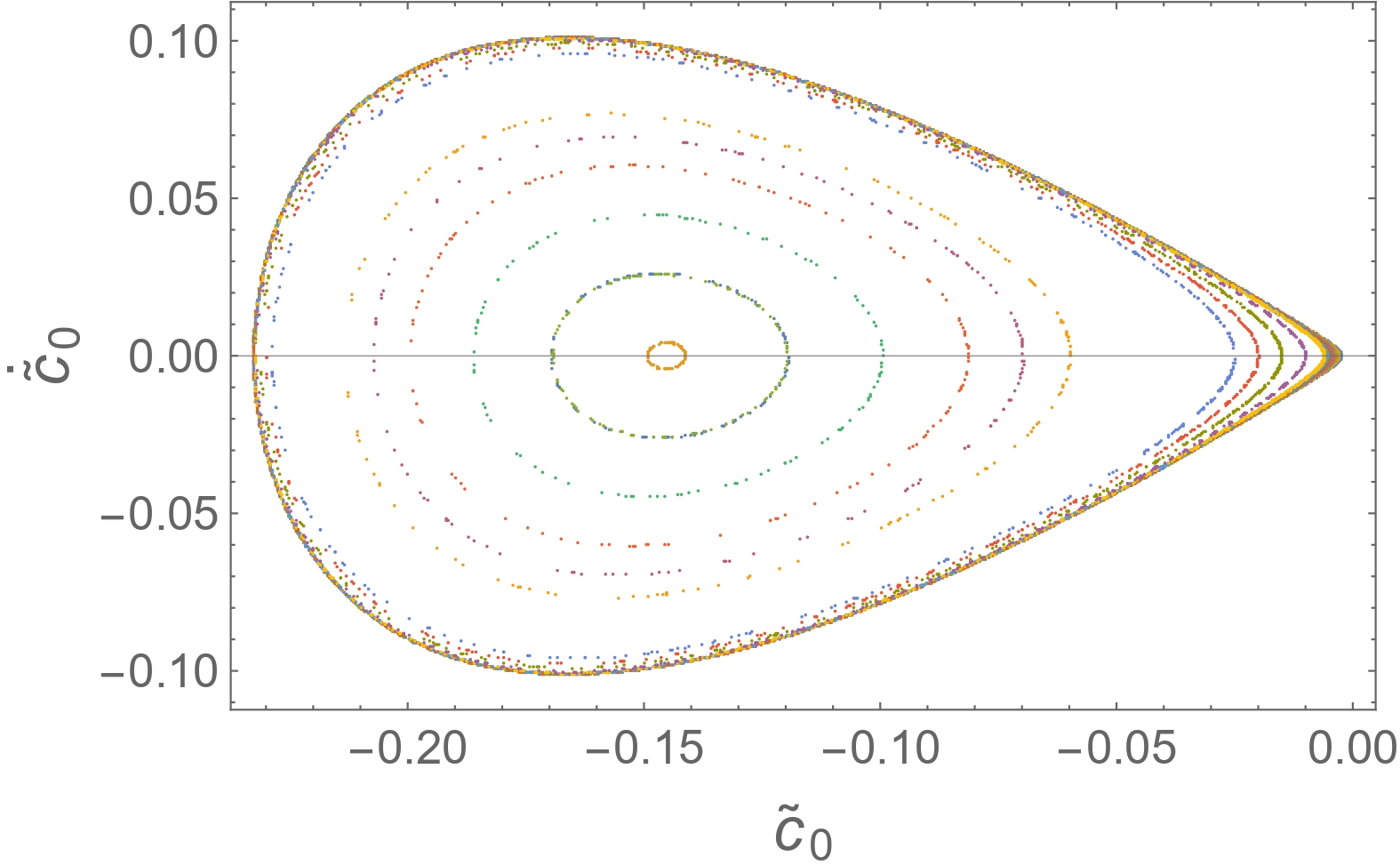} \quad
	 \includegraphics[width=0.4 \textwidth]{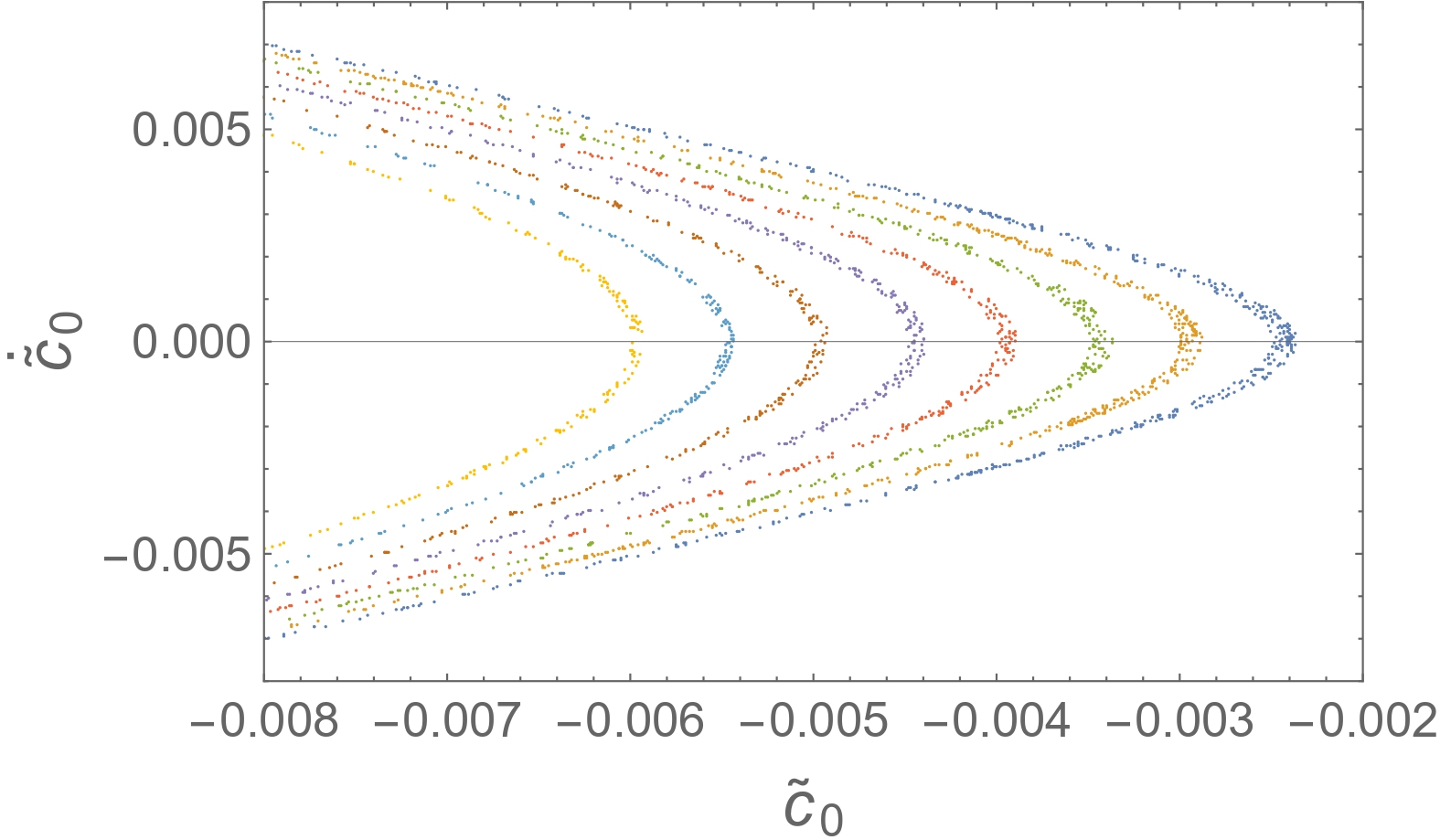}}}}  \quad
\makebox[\linewidth][c]{{
	{\includegraphics[width=0.4 \textwidth]{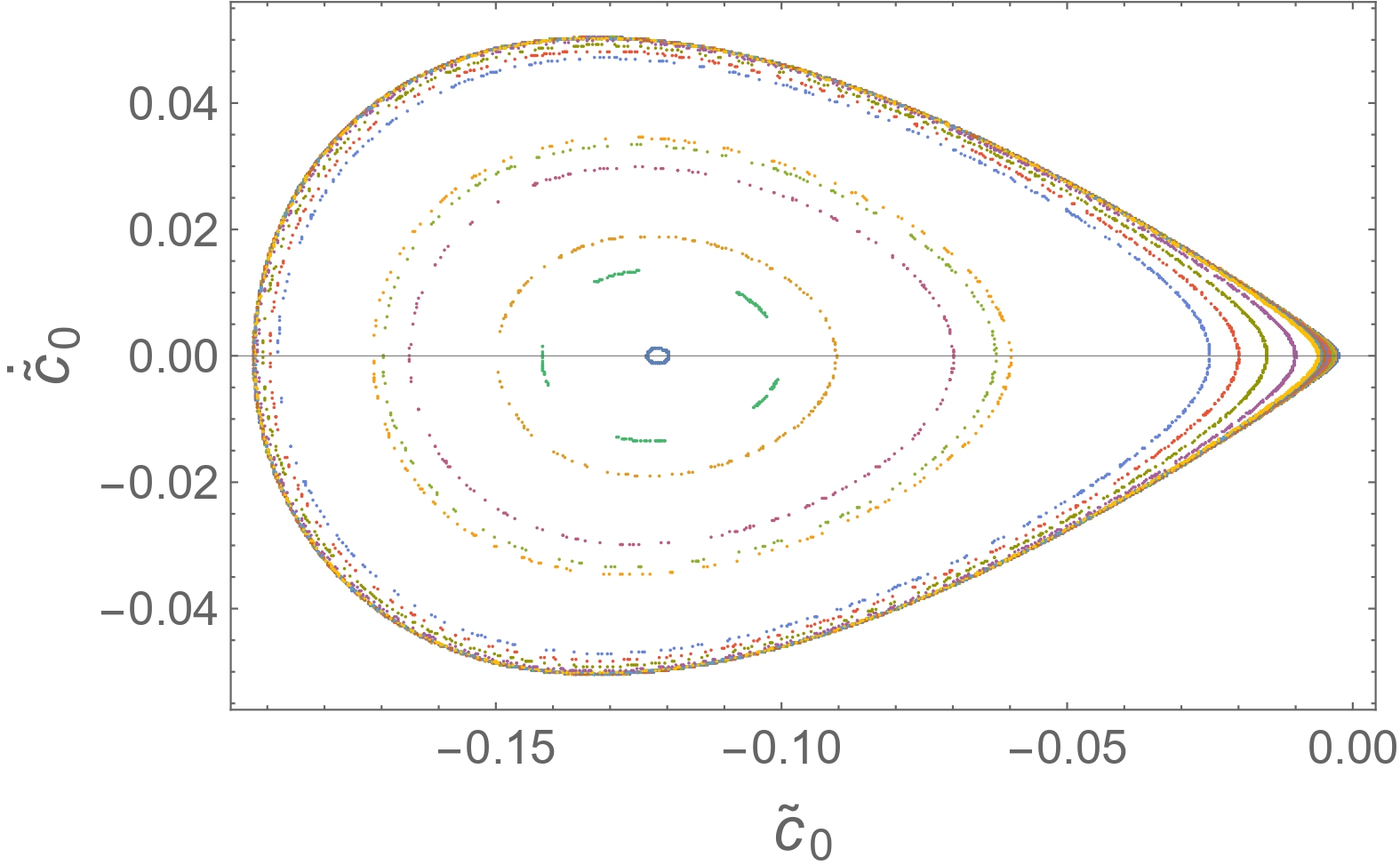}
\quad
	 \includegraphics[width=0.4 \textwidth]{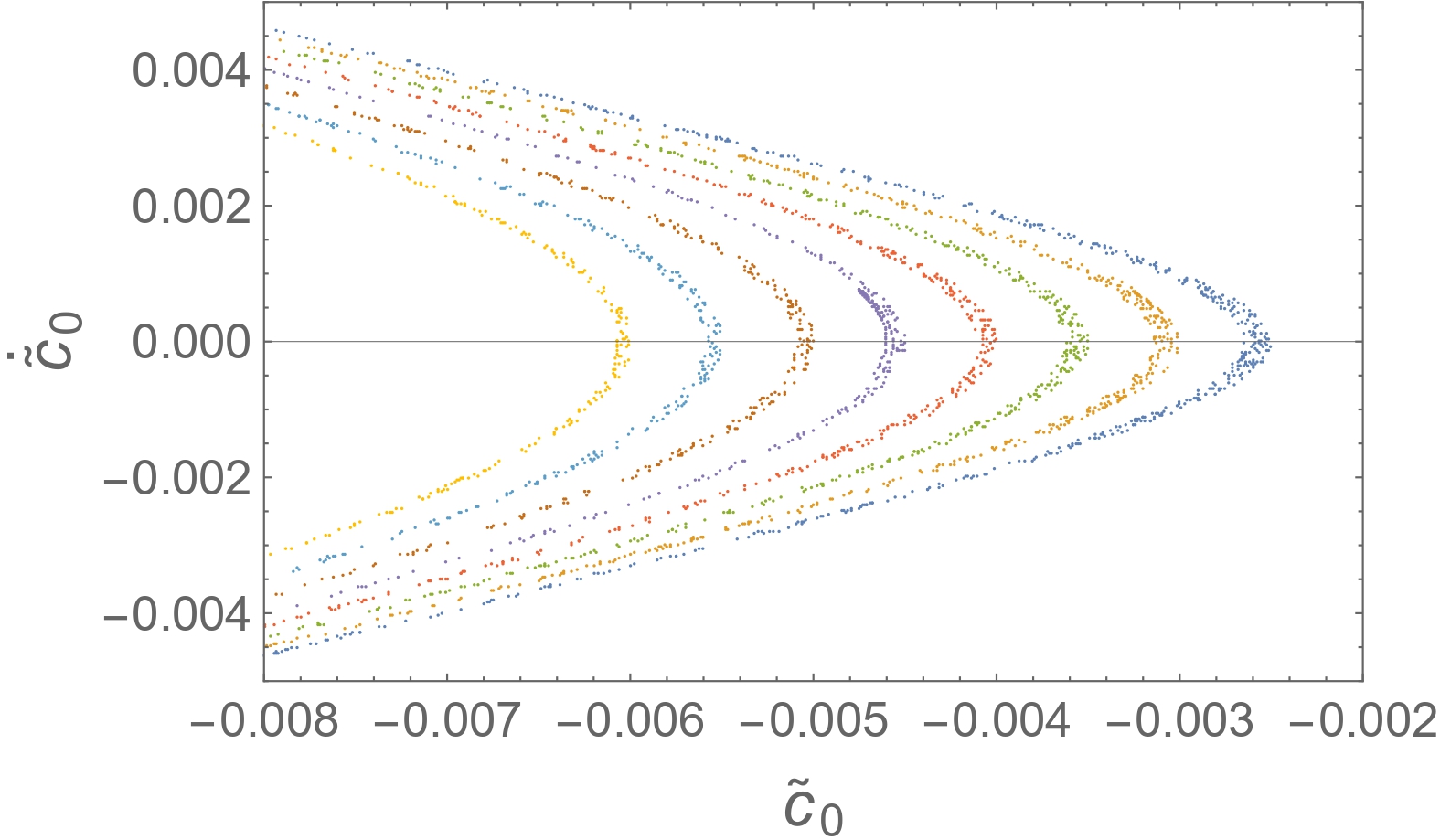}}}}
\quad
\caption[Poincar\'e plots for $r_0 = 1.1$ and different values of $B$]{\small Poincar\'e sections for a time-dependent perturbed string, obtained changing the initial conditions, with $r_0 = 1.1$ and increasing the chemical potential $\mu = 0.3$ (top row), $\mu= 0.6$ (middle row) and $\mu = 0.9$
(bottom row), for $\tilde{c}_1 = 0$ and $\dot{ \tilde{c}}_1 \ge 0$. The plots in the right column enlarge the corresponding ones in the left column in the range of small $\tilde{c}_0$, $\dot{\tilde{c}}_0$.}
\label{Fig:PoincarèMu}
\end{figure*}

\begin{figure*}[t!]
\centering
\makebox[\linewidth][c]{{
	{\includegraphics[width=0.4 \textwidth]{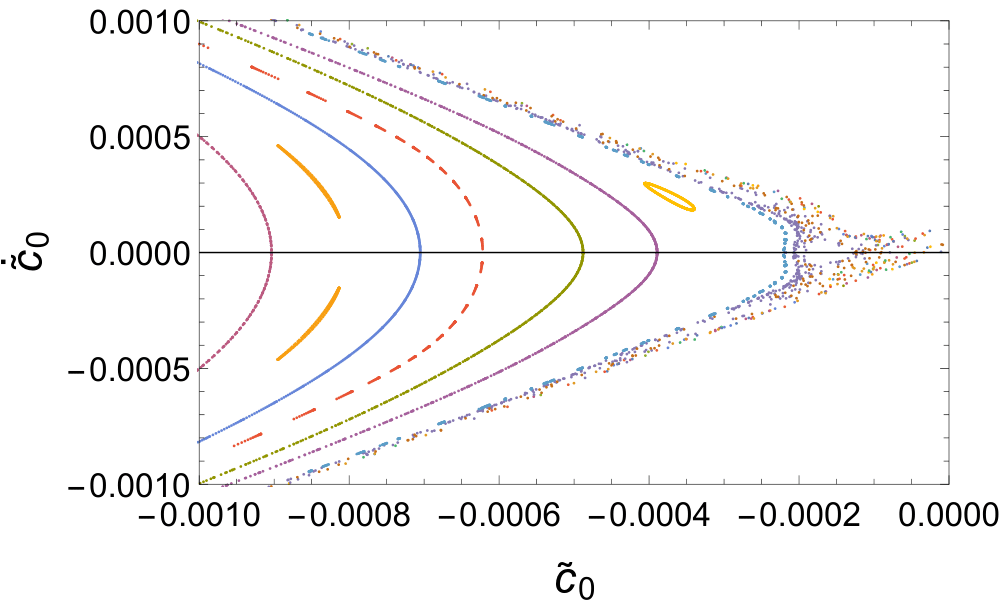} \quad
	 \includegraphics[width=0.4 \textwidth]{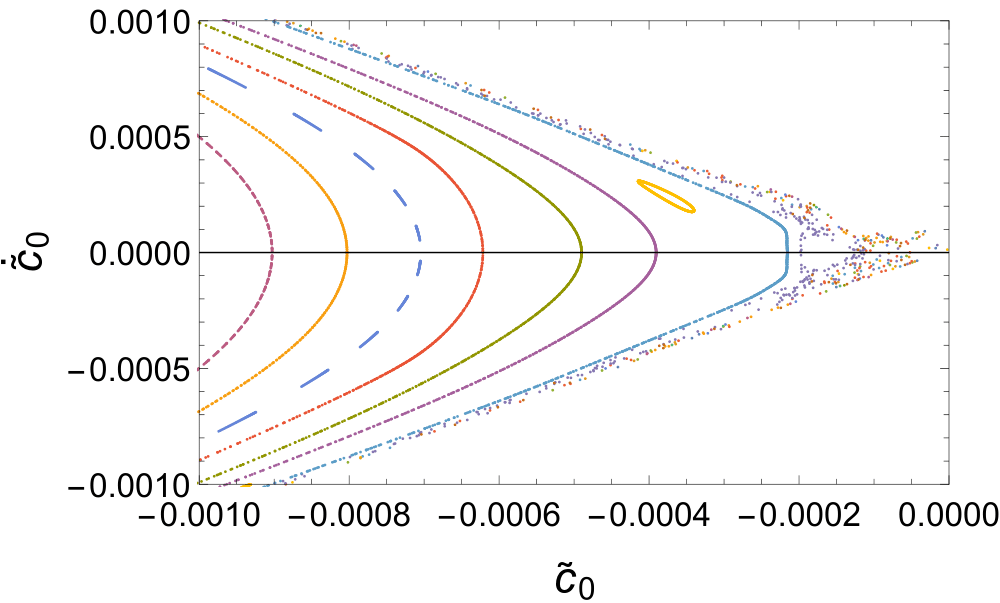}}}} \quad
\makebox[\linewidth][c]{{
	{\includegraphics[width=0.4 \textwidth]{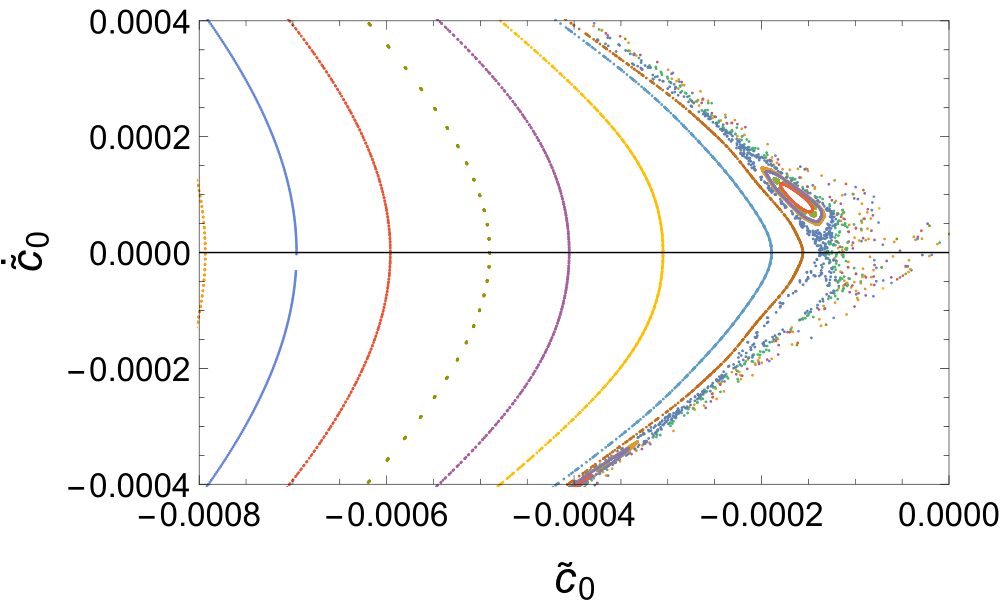} \quad
	 \includegraphics[width=0.4 \textwidth]{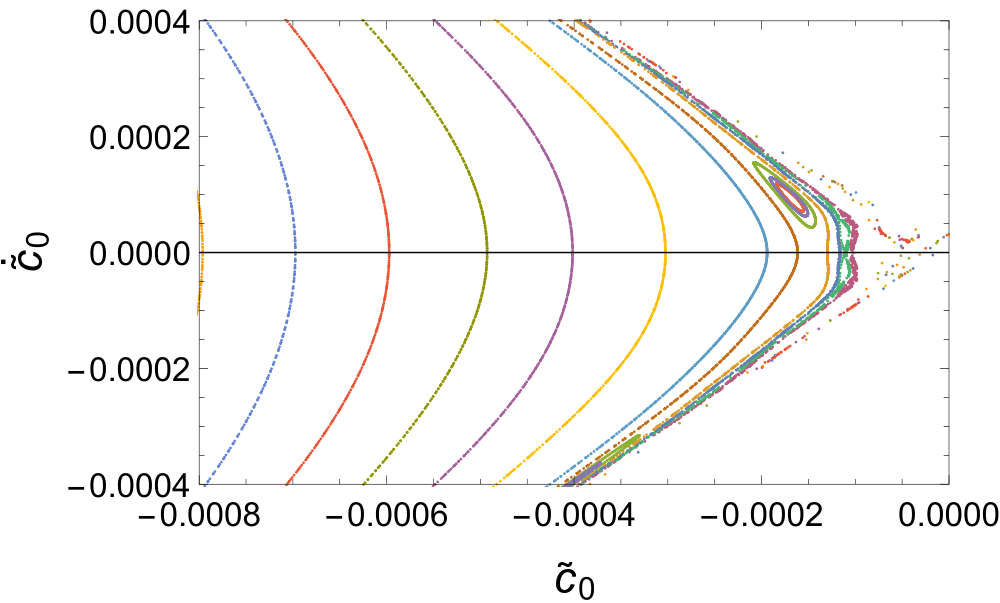}}}}  \quad
\makebox[\linewidth][c]{{
	{\includegraphics[width=0.4 \textwidth]{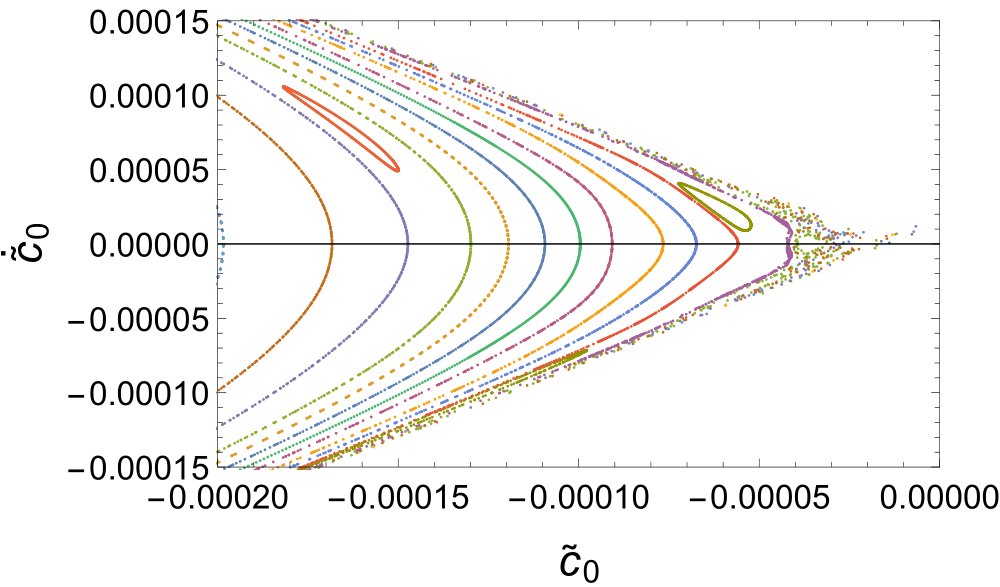}
\quad
	 \includegraphics[width=0.4 \textwidth]{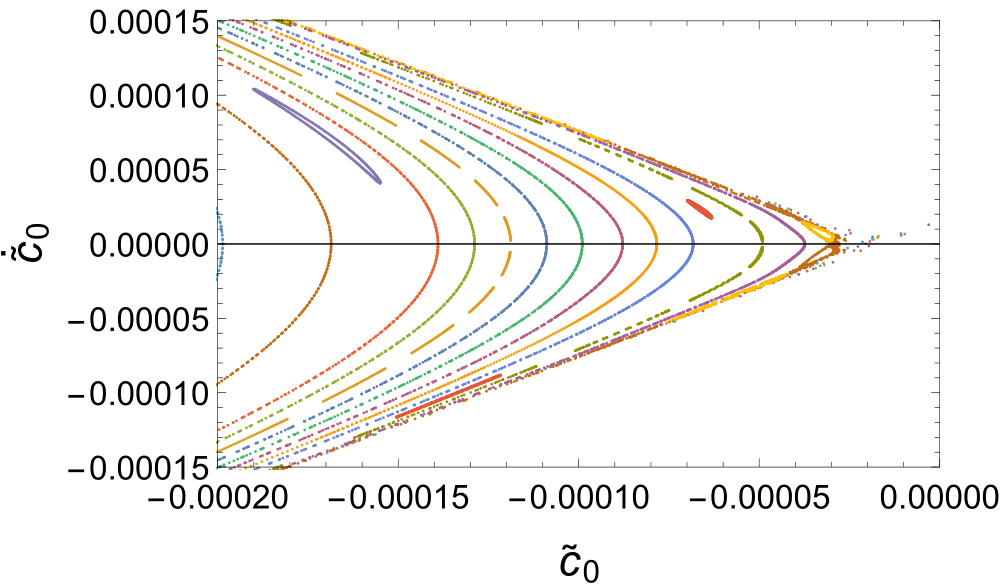}}}}
\quad
\caption[Poincar\'e plots for $r_0 = 1.1$ and different values of $B$]{\small Poincar\'e sections for a perturbed string in the  $x_1$ (left column) and $x_3$ configurations (right column). The initial conditions are changed with fixed energy $E = 10^{-5}$ and $r_0=1.1$. The magnetic field is increased from  $B=0.3$ (top row) to $B=0.6$ (middle row) and $B=1$ (bottom row). The sections  correspond to  $\tilde{c}_1 =0$ and $\dot{\tilde{c}}_1 \geqslant 0$.}
\label{Fig:Poincarè}
\end{figure*}

\noindent
We now perturb the static string, and from Eq.~\eqref{eq:13} we obtain the dynamics of the fluctuation through the functions $c_0 \left( t \right)$ and $c_1 \left( t \right)$.
The onset of chaos is displayed by the  Poincar\'e sections.
We construct the sections defined by $\tilde{c}_1 \left( t \right) = 0$ and $\dot{\tilde{c}}_1 \left( t \right)
\geqslant 0$ for bounded orbits within the trap in the potential. Setting $r_h=1$ and $r_0 = 1.1$ for both metrics, we obtain the sections Fig.~\ref{Fig:PoincarèMu},\ref{Fig:Poincarè} for different values of $\mu$ and $B$. For  $\tilde{c}_0$ near zero the orbits are scattered points which depend on the initial conditions. 
Increasing $\mu$ or $B$ the points in the sections arrange in more regular paths, showing that the effect of switching on the magnetic field, or the interaction with a baryon reservoir is to mitigate the chaotic behavior. 

\noindent
In Fig.~\ref{Fig:Poincarè} we observe that when the string is along $x_3$ we go closer to $\tilde{c}_0 = 0$ to observe chaos.
In the Poincar\'e plots we set $\tilde c_1 = 0$. In this case the potential has a trap for $\tilde c_0 < 0$. Considering the definition of the perturbation given in Eqs.~\eqref{eq:4} and  ~\eqref{eq:10}, the perturbation characterized by $\tilde c_1 = 0$ and $\tilde c_0 < 0$, corresponding to $c_1 = 0$ and $c_0 < 0$, describes a string moving away from the event horizon. Therefore, for $(\tilde c_0,\tilde c_1) = (0,0)$, hence for $(c_0,c_1) = (0,0)$, the tip of the string is closest to the horizon.

\begin{figure}[h!]
\centering
\makebox[\linewidth][c]{{
	{\includegraphics[width=0.5 \textwidth]{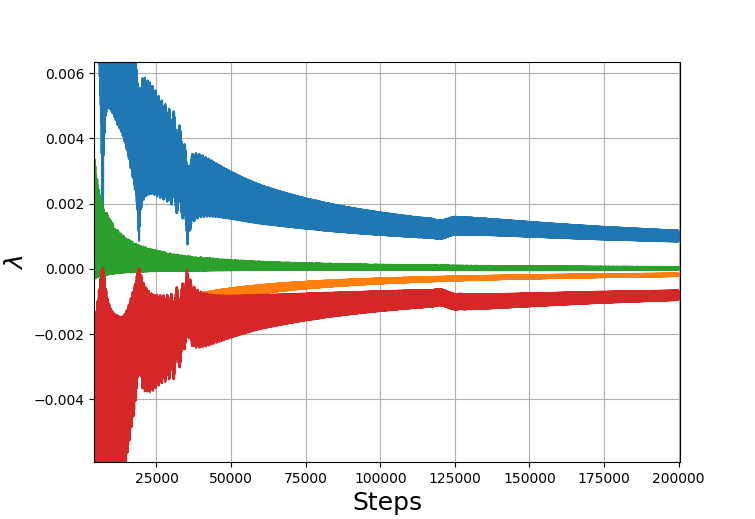} \quad
\includegraphics[width=0.5 \textwidth]{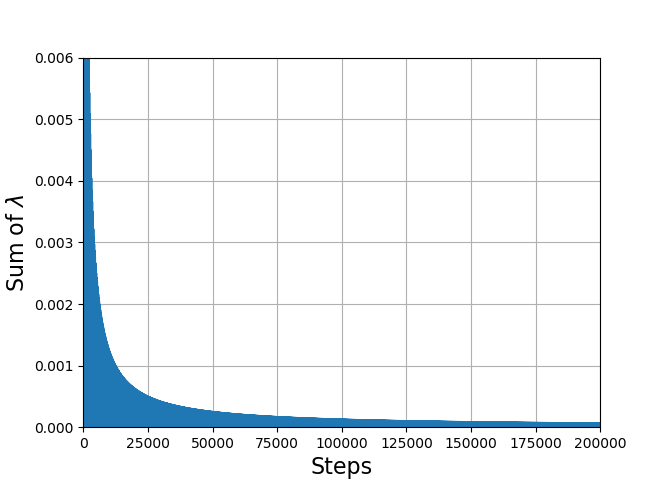}}}}
\caption[Convergency plots of the Lyapunov exponents for a string with $r_0=1.1$]{\baselineskip 12 pt \small Left: convergency plot of the four Lyapunov exponents for  a string along $x_1$, with $r_0=1.1$ and $B=0.6$. $2\times 10^5$ time steps are shown. For the initial conditions, the energy is set to $E = 10^{-5}$ together with  $\tilde{c}_0 = -0.0002$, $\dot{\tilde{c}}_0 = 0$, $\tilde{c}_1 = 0.0011$. Right: sum of the Lyapunov exponents for the same value of $B$.}
\label{Fig:XLCE}
\end{figure}

We  evaluate the Lyapunov exponents. In the four dimensional $c_0$, $c_1$ phase-space they can be computed for different values of $\mu$ or $B$ using the numerical method described in \cite{sandri1996numerical}. A convergency plot is shown in Fig.~\ref{Fig:XLCE}, together with the sum of the Lyapunov exponents which converges to zero with the evolution.

\begin{figure}[b!]
\centering
	\makebox[\linewidth][c]{{
	{\includegraphics[width=0.5 \textwidth]{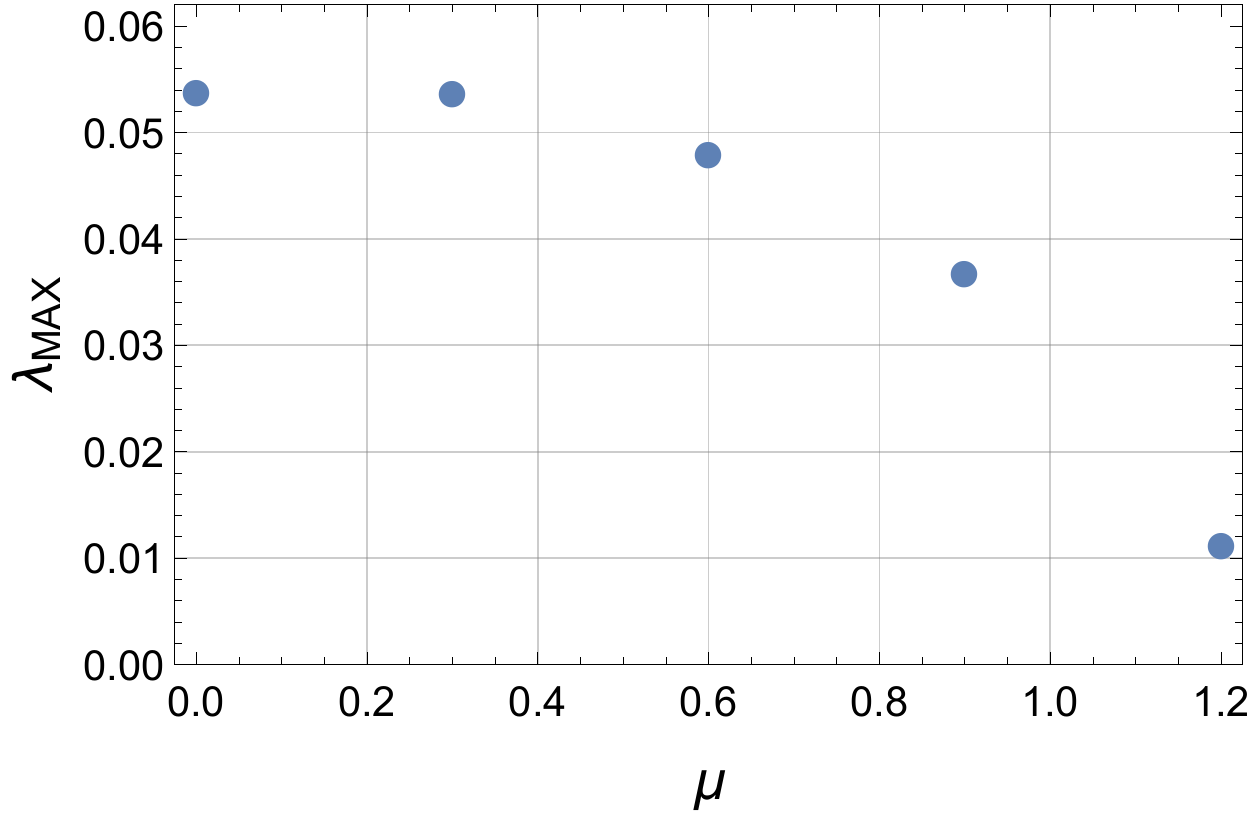} \quad
\includegraphics[width=0.54 \textwidth]{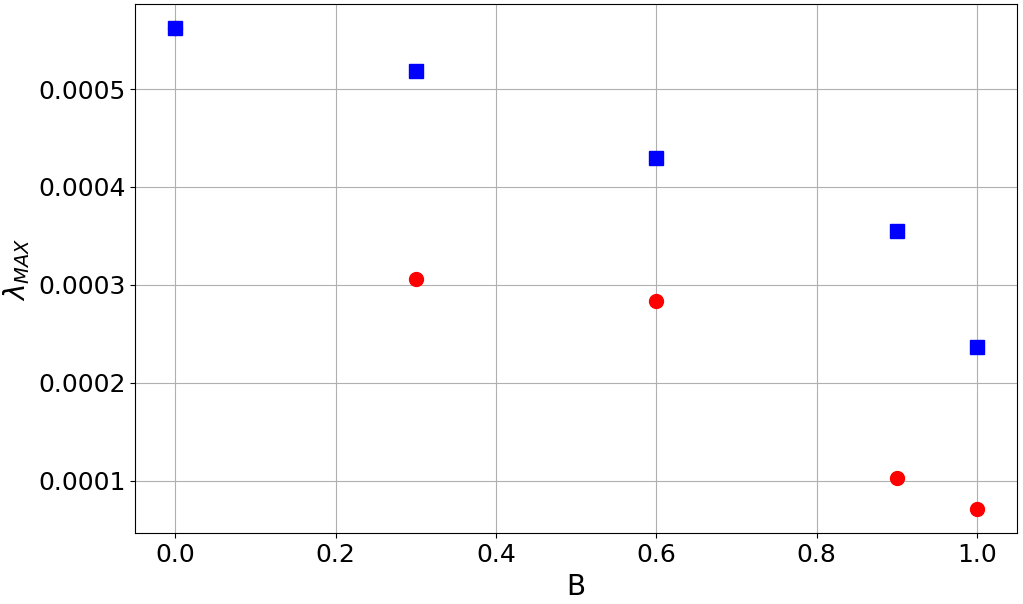}}}}
	\caption[Largest Lyapunov exponent as a function of $B$ for $x_1$ configuration]{\small Largest Lyapunov exponent  $\lambda_{MAX}$ versus $\mu$ (left) and $B$ (right) for $r_0=1.1$. In the plot at right the results for the  $x_1$  (blue squares) and $x_3$ string configurations (red points) are shown.}
\label{Fig:LCEmax}
\end{figure}

\noindent
The convergency plot is a damped oscillating function. The value of the largest Lyapunov exponent can be extrapolated fitting the maximum in each oscillation  and considering  $t \to + \infty$.
The values obtained decrease as $\mu$ or $B$ increases, as shown in Fig.~\ref{Fig:LCEmax}: the effect 
of the magnetic field and the chemical potential is to soften the dependence on the initial conditions, making the string less chaotic. 
\noindent
At the right in Fig.~\ref{Fig:LCEmax}  the results for the two configurations are compared.
For the same  values of $B$ and $r_0$, so at the same distance from the BH horizon, smaller Lyapunov exponents are found in the $x_3$ configuration. 
The Poincar\'e plots show that chaos is produced in the proximity of the BH horizon, and that  the string dynamics  is less chaotic if the chemical potential or the magnetic field increases. This is confirmed by the largest Lyapunov exponent.
As we can see from the plots in Fig.~\ref{Fig:LCEmax}, the MSS bound is satisfied for both the metrics, since $5/3 \le \lambda_{MSS} \le 2$ and $0.56 \le \lambda_{MSS} \le 2$ for the two considered cases.

The AdS-RN metric in Eq.~\eqref{eq:RNLine} can be modified with a warp factor, used to implement a confinement mechanism in holographic models of QCD \cite{Colangelo:2013ila}. The line element is defined as

\begin{equation}
d s^2 = e^{-\frac{c^2}{r^2}} \left(- f \left( r \right) r^2 d t^2 + r^2 d \vect{x}^2 +\frac{1}{ r^2 f \left( r \right)} d r^2 \right) ,
\label{eq:DilatonRNLine}
\end{equation} 
\noindent
with metric function $f(r)$ in \eqref{eq:RNMF}. The Hawking temperature does not depend on the dilaton parameter $c$, therefore it is given in Eq.~\eqref{eq:Tvsrhmu}. The warp factor mainly affects the IR small r region, while the geometry became asymptotically AdS$_5$ in the UV $r \rightarrow \infty$ region. A dilaton factor has been used to study features of the QCD phenomenology at finite temperature and baryon density, namely the behaviour of the quark and gluon condensates increasing $T$ and $\mu$, the phase diagram, and the in-medium broadening of the spectral functions of two-point correlators \cite{Colangelo:2010pe,Colangelo:2011sr,Colangelo:2012jy,Colangelo:2013ila}

\begin{figure}[h!]
\centering
	\makebox[\linewidth][c]{{
	{\includegraphics[width=0.5 \textwidth]{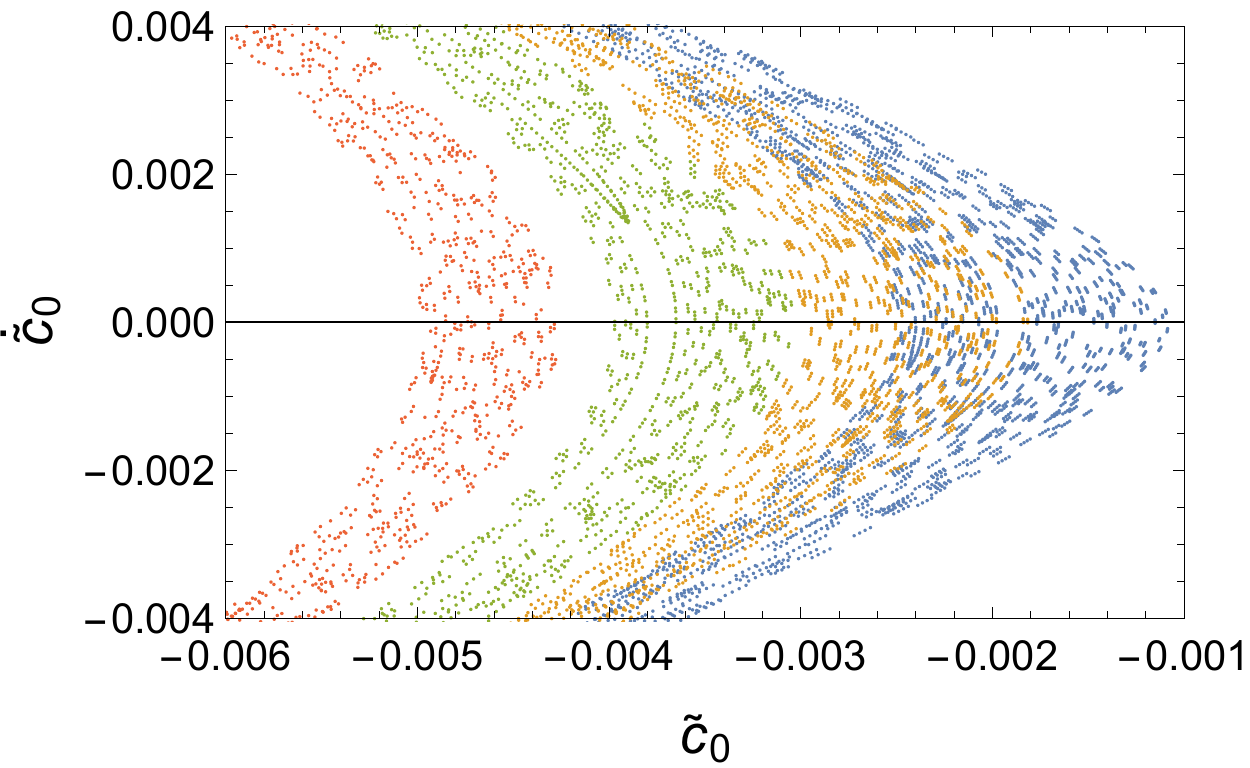} \quad
\includegraphics[width=0.5 \textwidth]{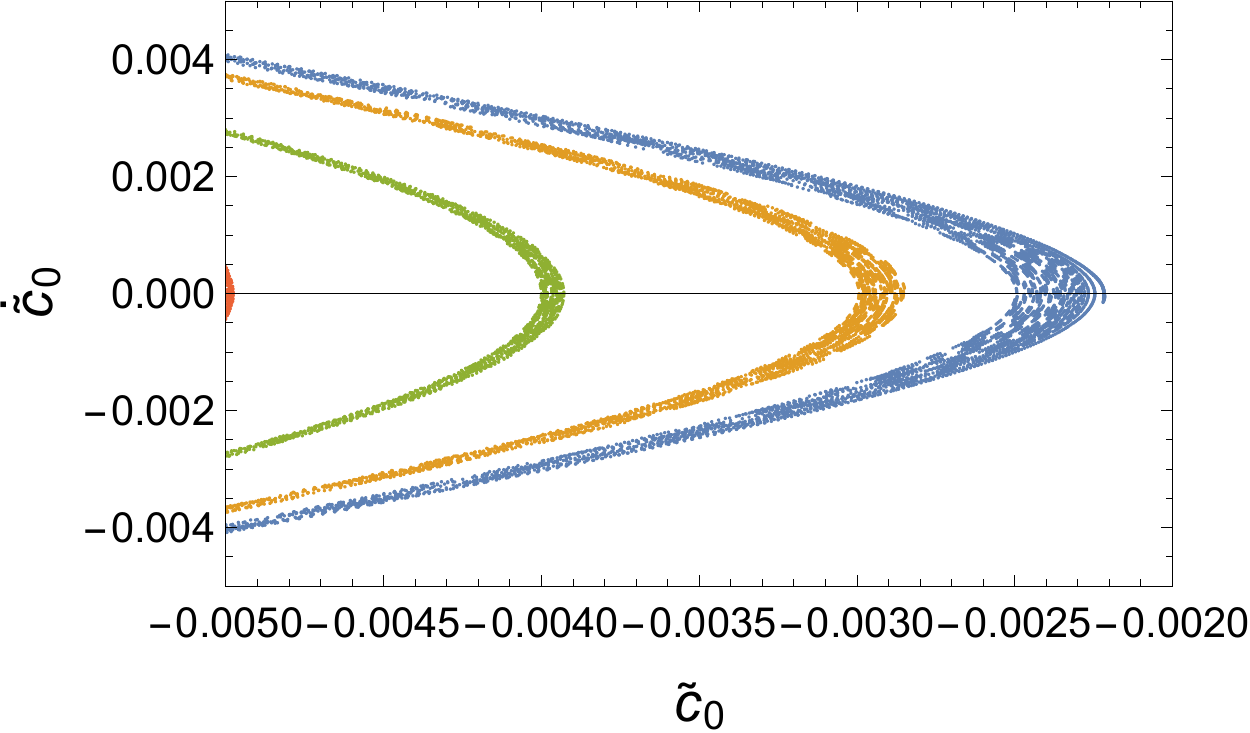}}}}
	\caption[Largest Lyapunov exponent as a function of $B$ for $x_1$ configuration]{\small Zoom in the small $\tilde{c}_0$, $\dot{\tilde{c}}_0$ region of the poincar\'e section for the perturbed string in the background geometry with warp factor \eqref{eq:DilatonRNLine}, for $r_0 = 1.1$, $\mu = 0$ andparameter of the dilaton $c=1$ (left) and $c=2$ (right). The sections are obtained setting the energy $E = 1 \times 10^{-5}$ and a time evolution of $8 \times 10^{-3}$ time steps.}
\label{Fig:DilatonPoincare}
\end{figure}

To study the dependence on the dilaton parameter $c$, we inspect the Poincar\'e plots and compute the Lyapunov exponents. The Poincar\'e sections for $r_h = 1$, $r_0 = 1.1$, $\mu = 0$ at different $c$ are shown in Fig.~\ref{Fig:DilatonPoincare}.

\begin{figure}[h!]
\begin{center}
\includegraphics[width=0.5 \textwidth]{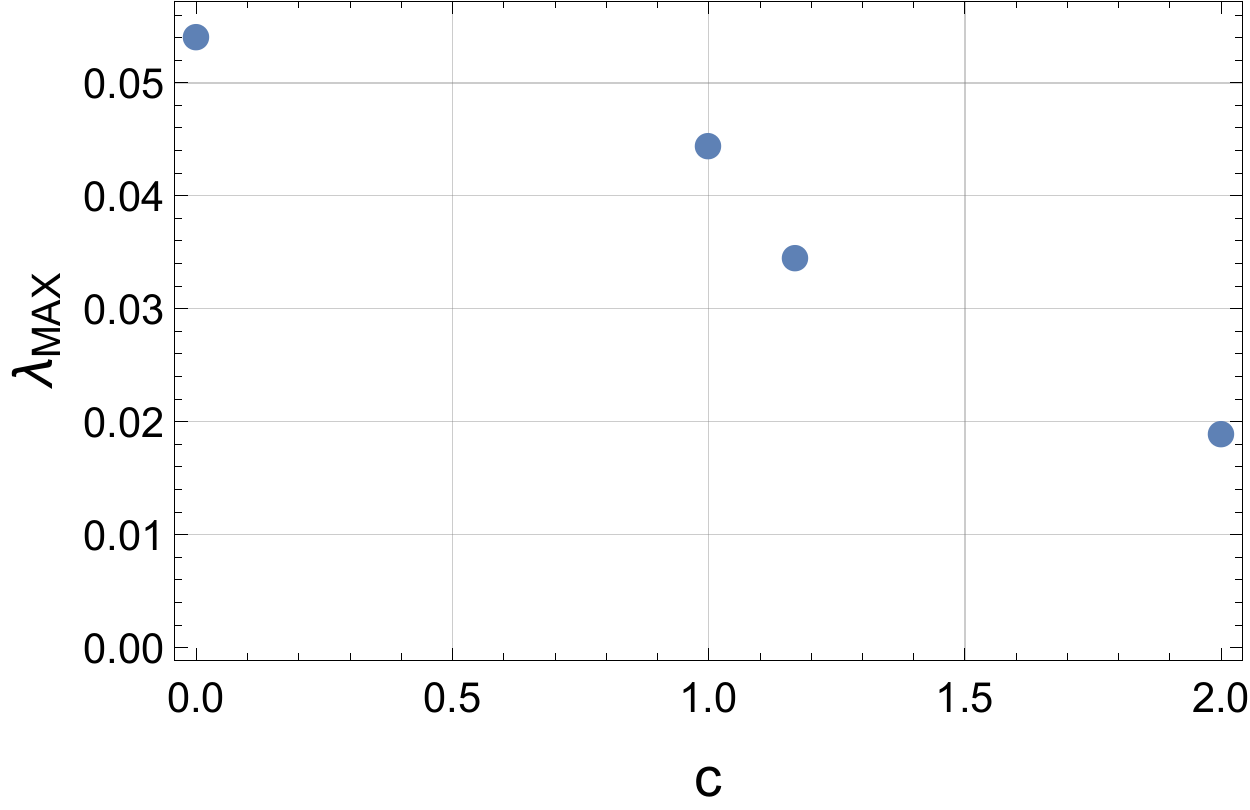}
\caption{\small {Largest Lyapunov exponent for $r_0 = 1.1$ and $\mu=0$, increasing the dilaton constant.}}
\label{Fig:DilatonLyapunov}
\end{center}
\end{figure}

\noindent
As we can see from the Poincar\'e section increasing the dilaton parameter stabilizes the system. This is confirmed by the maximum Lyapunov exponent as a function of $c$, Fig~\ref{Fig:DilatonLyapunov}.

\noindent
The MSS bound is satisfied, since $\lambda_{MAX} \ll \lambda_{MSS} = 2$ for all values of the dilaton parameter $c$.

\section{Conclusions}\label{conclusions}

We have presented a method to explore the chaotic behaviour of a strongly coupled $Q \bar Q$ system at finite temperature through its gravity dual system. This allow us to test the Maldacena, Shenker and Stanford conjecture. Chaos has been observed in the Poincar\'e plots, characterized by scattered points in the region close to the black hole horizon, and quantitatively described computing the Lyapunov exponents. The system becomes less chaotic increasing  $\mu$, $B$ and $c$. For the magnetic field case, anisotropy effect in two different orientations of the string is found. The stabilization effect of the magnetic field is stronger for the configuration with string endpoints lying on a line parallel to the field. The MSS bound \eqref{eq:1} is satisfied for the largest Lyapunov exponent and therefore, also the generalization \eqref{eq:2}. The MSS bound remains universal.

\vspace*{1.cm}
\noindent {\bf Acknowledgements.}
I thank P. Colangelo, F. De Fazio and F. Giannuzzi, co-authors of the works on which this review is based on.
This study has been carried out within the INFN project (Iniziativa Specifica) QFT-HEP.

\bibliographystyle{JHEP}
\bibliography{ReviewNL}

\end{document}